\documentclass[twocolumn,showpacs,prx,superscriptaddress,floatfix]{revtex4}

\usepackage{color}
\usepackage{tabularx}
\usepackage{epsfig}
\usepackage{amsmath}
\usepackage{amssymb}
\usepackage{graphicx}
\usepackage{wasysym}
\usepackage{dcolumn}
\usepackage{bm}
\usepackage{subfigure}
\usepackage{psfrag}
\usepackage{subfigure}

\begin{document}

\title{Glassy Chimeras Could Be Blind to Quantum Speedup: \\ Designing
Better Benchmarks for Quantum Annealing Machines}

\author{Helmut G.~Katzgraber}
\affiliation{Department of Physics and Astronomy, Texas A\&M University,
College Station, Texas 77843-4242, USA}
\affiliation{Materials Science and Engineering Program, Texas A\&M
University, College Station, Texas 77843, USA}

\author{Firas Hamze}
\affiliation {D-Wave Systems, Inc., 3033 Beta Avenue, Burnaby, British
Columbia, V5G 4M9, Canada}

\author{Ruben S.~Andrist}
\affiliation{Santa Fe Institute, 1399 Hyde Park Road, Santa Fe New
Mexico 87501, USA}

\date{\today}

\begin{abstract}

Recently, a programmable quantum annealing machine has been built that
minimizes the cost function of hard optimization problems by, in
principle, adiabatically quenching quantum fluctuations. Tests performed
by different research teams have shown that, indeed, the machine seems
to exploit quantum effects.  However experiments on a class of
random-bond instances have not yet demonstrated an advantage over
classical optimization algorithms on traditional computer hardware.
Here we present evidence as to why this might be the case. These
engineered quantum annealing machines effectively operate coupled to a
decohering thermal bath. Therefore, we study the finite-temperature
critical behavior of the standard benchmark problem used to assess the
computational capabilities of these complex machines. We simulate both
random-bond Ising models and spin glasses with bimodal and Gaussian
disorder on the D-Wave Chimera topology.  Our results show that while
the worst-case complexity of finding a ground state of an Ising spin
glass on the Chimera graph is not polynomial, the finite-temperature
phase space is likely rather simple because spin glasses on Chimera have
only a zero-temperature transition. This means that benchmarking
optimization methods using spin glasses on the Chimera graph might not
be the best benchmark problems to test quantum speedup.  We propose
alternative benchmarks by embedding potentially harder problems on the
Chimera topology. Finally, we also study the (reentrant)
disorder-temperature phase diagram of the random-bond Ising model on
the Chimera graph and show that a finite-temperature ferromagnetic phase
is stable up to 19.85(15)\% antiferromagnetic bonds.  Beyond this
threshold, the system only displays a zero-temperature spin-glass phase.
Our results therefore show that a careful design of the hardware
architecture and benchmark problems is key when building quantum
annealing machines.

\end{abstract}


\maketitle

\section{Introduction}

Quantum devices are gaining an increasing importance in everyday
technology: They find applications in different technological areas such
as (true) quantum random number generators, as well as quantum
encryption systems for data transmission. The holy grail is to build a
programmable quantum simulator with capabilities exceeding
``traditional'' computer hardware based on classical bits. The first
programmable commercial device that attempts to exploit the unique power
of quantum mechanics to perform computations is the D-Wave One quantum
annealer \cite{comment:d-wave}. In analogy to simulated annealing
\cite{kirkpatrick:83} where thermal fluctuations are adiabatically
quenched to minimize a cost function, this machine is based on the
quantum annealing optimization method
\cite{finnila:94,kadowaki:98,kadowaki:98a,brooke:99,martonak:02,santoro:02,santoro:06,das:05,das:08}
where quantum fluctuations replace thermal ones.

\begin{figure}[h]
\includegraphics[width=0.65\columnwidth]{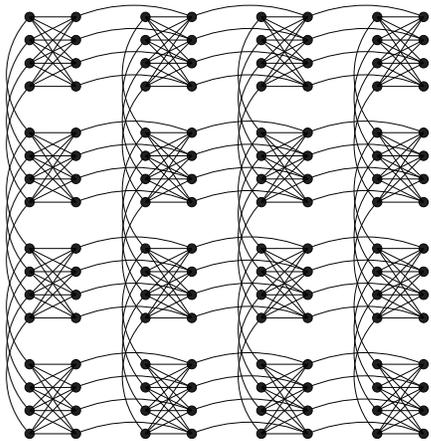}
\caption{
Example Chimera graph with $k^2 = 16$ blocks of $8$ qubits (black dots).
This means the system has $N = 128$ qubits and an effective linear size
$L = \sqrt{N} = \sqrt{128}$. The high connectivity between the spins
within each block effectively renders the model quasi-two dimensional.
Note that the graph is not planar.
}
\label{fig:chimera}
\end{figure}

Tests by different research teams suggest that, indeed, the D-Wave
quantum annealer likely optimizes using quantum effects
\cite{johnson:11,boixo:13,boixo:13a,wang:13,dickson:13}.  Although it
has been shown theoretically \cite{morita:08}, as well as with numerical
experiments \cite{santoro:02,comment:santoro} that quantum annealing
could, in principle, outperform classical (thermal) optimization
algorithms (such as simulated annealing \cite{kirkpatrick:83}) on an
algorithmic level, when applied to a class of random edge-weight
instances, the quantum annealing machine has not yet shown a speedup
over classical optimization methods \cite{boixo:13,ronnow:14}.  In this
work, we present evidence for why this might be the case: The D-Wave One
and Two quantum annealing machines use a restrictive ``Chimera''
topology (see Fig.~\ref{fig:chimera} for an example with 128 quantum
bits) imposed because of fabrication constraints of the solid-state
quantum bits.

Probably the best benchmark problem to test the efficiency of
optimization algorithms is a spin glass \cite{binder:86,lucas:14}. Both
the disorder and frustration produce a complex energy landscape that
challenges optimization algorithms.  As such, all current benchmarks of
the quantum annealing machine attempt to find the ground state of a
certain class of Ising spin glass on the Chimera topology.  However, as
shown in this work, instances belonging to this class of Ising spin
glasses on the Chimera topology only have a spin-glass phase at zero
temperature. Furthermore, the energy landscape of such problems seems to
be simpler down to low temperatures than for a system with a
finite-temperature transition because correlations only build up very
close to absolute zero.  Because quantum annealing excels in tunneling
through barriers---barriers that do not seem to be very pronounced at
finite, but low temperatures in this case---classical annealing
schedules might typically have an advantage for this particular class of
systems.

Although the worst-case complexity of finding a ground state of an Ising
spin glass on the Chimera topology is worse than polynomial
\cite{barahona:82}, it seems that the fact that the system only orders
at zero temperature allows for an easier determination of {\em typical}
ground-state instances using heuristic classical approaches
\cite{boixo:13,comment:tail}. As such, Ising spin glasses on the Chimera
topology live up to their name: an amalgamation of both ordinary and
complex behavior.

We reach this conclusion by studying the critical behavior of Ising spin
glasses with both Gaussian and bimodal random bonds on the Chimera
topology, as well as the random-bond Ising model. Based on our findings,
we propose stronger benchmark problems by embedding on the Chimera
topology problems that should have a finite-temperature transition and
thus might be harder to optimize. Furthermore, our results show that a
careful design of the hardware architecture and benchmark problems is
key when building quantum annealing machines.

We should also mention that while the results of this paper provide a
plausible explanation for the scaling behavior observed so far on
random-bond Ising problems, it is also known that on quantum annealer
implementations, the couplers and biases are influenced by various
sources of noise and error, as demonstrated by the fact that
gauge-transformed specifications of the same problem can give
substantially different performance \cite{boixo:13}. Classical simulated
annealing does not, of course, have this issue, and it is currently
unclear how much loss of efficiency these errors cause for the hardware.

The paper is structured as follows. In Sec.~\ref{sec:model} we introduce
the standard benchmark model. Results within the spin-glass sector are
presented in Sec.~\ref{sec:results_sg}, followed by results within the
ferromagnetic sector in Sec.~\ref{sec:results_fm}.  In
Sec.~\ref{sec:discussion}, we discuss better benchmarks, followed by
concluding remarks.

\section{Model, Observables, and Algorithm}
\label{sec:model}

We study the spin-glass Hamiltonian 
\begin{equation}
{\mathcal H} = - \sum_{i,j = 1}^N J_{ij} S_i S_j , 
\label{eq:ham}
\end{equation}
with $S_i \in \{\pm 1\}$ Ising spins on the nodes of
the Chimera graph. An example of the topology with $4 \times 4$ blocks
of $8$ spins is shown in Fig.~\ref{fig:chimera}. A chimera graph with $k
\times k$ blocks has $N = 8k^2$ spins and a characteristic linear length
scale of $L = \sqrt{N}$. The interactions $J_{ij}$ are either chosen
from a Gaussian disorder distribution with zero mean and unit variance
or from a bimodal distribution ${\mathcal P}(J_{ij}) = p\delta(J_{ij} -
1) + (1-p)\delta(J_{ij} + 1)$, where with a probability $p$ a bond is
ferromagnetic.

Ordering in spin glasses is detected from the spin overlap $q=(1/N)
\sum_i S^{\alpha}_i S^{\beta}_i$, where ``$\alpha$'' and ``$\beta$'' are
two independent spin replicas with the same disorder.  In the
ferromagnetic case, order is measured via the magnetization, i.e.,
$m=(1/N) \sum_i S^{\alpha}_i$.  To detect the existence of a phase
transition to high precision, we measure the Binder ratio
\cite{binder:81} $g_{\mathcal O} = (1/2)[3 - \langle {\mathcal O}^4
\rangle/\langle {\mathcal O}^2\rangle^2]$, where ${\mathcal O}$
represents either the magnetization $m$ for the ferromagnetic sector or
the spin overlap $q$ in the spin-glass sector. The Binder ratio is a
dimensionless function, which means that, at a putative transition, data
for different characteristic system sizes $L$ will cross when $T = T_c$,
where $T_c$ is the critical temperature (up to corrections to scaling).
This means $g_{\mathcal O} \sim G_{\mathcal O}[L^{1/\nu_{\mathcal O}}(T
- T_c^{\mathcal O})]$.  Using a finite-size scaling analysis, the
critical temperature $T_c^{\mathcal O}$ and the critical exponent
$\nu_{\mathcal O}$ can be determined. To uniquely determine the
universality class of a system, two critical exponents are needed
\cite{yeomans:92}. To this end, we also measure the susceptibility
$\chi_{\mathcal O} = N \langle {\mathcal O}^2 \rangle$, where ${\mathcal
O}$ again represents either the magnetization $m$ for the ferromagnetic
sector, or the spin-glass order parameter $q$ for the spin-glass sector.
The susceptibility scales as $\chi_{\mathcal O} \sim L^{2 -
\eta_{\mathcal O}} C_{\mathcal O}[L^{1/\nu_{\mathcal O}}(T -
T_c^{\mathcal O})]$, where $\eta_{\mathcal O}$ is an independent
critical exponent.

Simulations are done using the replica exchange Monte Carlo
\cite{hukushima:96} method, and simulation parameters are listed in
Table \ref{tab:simparams}. Note that for each disorder instance we
simulate two independent replicas to compute the spin-glass overlap. In
the Gaussian case, we test equilibration using the method developed in
Ref.~\cite{katzgraber:01} adapted to the Chimera topology. For bimodal
disorder, we perform a logarithmic binning of the data. Once the last
four bins agree within error bars, we deem the system to be in thermal
equilibrium.  To obtain optimal values for the critical parameters, we
determine these via the analysis method pioneered in
Ref.~\cite{katzgraber:06}, where the critical parameters are optimized
using a Levenberg-Marquard minimization until the chi square of a fit to
a third-order polynomial is minimal. This approach is then bootstrapped
to obtain statistically sound error bars.

\begin{table}
\caption{
Simulation parameters: For each number of spins $N$ and fraction of
ferromagnetic bonds $p$, we equilibrate and measure for $2^b$ Monte Carlo
sweeps.  $T_{\rm min}$ [$T_{\rm max}$] is the lowest [highest]
temperature, and $N_T$ is the number of temperatures.  $N_{\rm sa}$ is
the number of disorder samples. The bottom block labeled with ``Gauss''
lists the simulation parameters for the pure spin glass with Gaussian
disorder. The numbers for the pure spin glass with bimodal disorder are
labeled with ``$p = 0.500$.'' 
\label{tab:simparams}}
\begin{tabular*}{\columnwidth}{@{\extracolsep{\fill}} l l c c c c c}
\hline
\hline
$p$ & $N$ & $b$ & $T_{\rm min}$ & $T_{\rm max}$ & $N_{T}$ & $N_{\rm sa}$ \\
\hline
$0.000$ 		  & $800$ &$21$ &$2.500$ &$5.500$ &$31$ & $128$ \\
$0.000$ 		  &$1152$ &$21$ &$2.500$ &$5.500$ &$31$ & $128$ \\
$0.000$ 		  &$1568$ &$21$ &$2.500$ &$5.500$ &$31$ & $128$ \\
$0.000$ 		  &$2048$ &$21$ &$2.500$ &$5.500$ &$31$ & $128$ \\
$0.000$ 		  &$2592$ &$21$ &$2.500$ &$5.500$ &$31$ & $128$ \\
$0.000$ 		  &$3200$ &$21$ &$2.500$ &$5.500$ &$31$ & $128$ \\
\\[-0.75em]
$0.040$, $0.080$, $0.120$ & $800$ &$21$ &$2.500$ &$5.500$ &$31$ &$2800$ \\
$0.040$, $0.080$, $0.120$ &$1152$ &$21$ &$2.500$ &$5.500$ &$31$ &$2800$ \\
$0.040$, $0.080$, $0.120$ &$1568$ &$21$ &$2.500$ &$5.500$ &$31$ &$2800$ \\
$0.040$, $0.080$, $0.120$ &$2048$ &$21$ &$2.500$ &$5.500$ &$31$ &$2800$ \\
$0.040$, $0.080$, $0.120$ &$2592$ &$21$ &$2.500$ &$5.500$ &$31$ &$2800$ \\
$0.040$, $0.080$, $0.120$ &$3200$ &$21$ &$2.500$ &$5.500$ &$31$ &$2800$ \\
\\[-0.75em]
$0.160$, $0.180$, $0.190$ & $800$ &$21$ &$1.500$ &$4.500$ &$31$ &$2800$ \\
$0.160$, $0.180$, $0.190$ &$1152$ &$21$ &$1.500$ &$4.500$ &$31$ &$2800$ \\
$0.160$, $0.180$, $0.190$ &$1568$ &$21$ &$1.500$ &$4.500$ &$31$ &$2800$ \\
$0.160$, $0.180$, $0.190$ &$2048$ &$21$ &$1.500$ &$4.500$ &$31$ &$2800$ \\
$0.160$, $0.180$, $0.190$ &$2592$ &$21$ &$1.500$ &$4.500$ &$31$ &$2800$ \\
$0.160$, $0.180$, $0.190$ &$3200$ &$21$ &$1.500$ &$4.500$ &$31$ &$2800$ \\
\\[-0.75em]
$0.195$, $0.197$          & $800$ &$21$ &$0.500$ &$3.500$ &$31$ &$2800$ \\
$0.195$, $0.197$          &$1152$ &$21$ &$0.500$ &$3.500$ &$31$ &$2800$ \\
$0.195$, $0.197$          &$1568$ &$21$ &$0.500$ &$3.500$ &$31$ &$2800$ \\
$0.195$, $0.197$          &$2048$ &$21$ &$0.500$ &$3.500$ &$31$ &$2800$ \\
$0.195$, $0.197$          &$2592$ &$21$ &$0.500$ &$3.500$ &$31$ &$2800$ \\
$0.195$, $0.197$          &$3200$ &$21$ &$0.500$ &$3.500$ &$31$ &$2800$ \\
\\[-0.75em]
$0.200$, $0.220$          & $800$ &$21$ &$1.500$ &$4.500$ &$31$ &$2800$ \\
$0.200$, $0.220$          &$1152$ &$21$ &$1.500$ &$4.500$ &$31$ &$2800$ \\
$0.200$, $0.220$          &$1568$ &$21$ &$1.500$ &$4.500$ &$31$ &$2800$ \\
$0.200$, $0.220$          &$2048$ &$21$ &$1.500$ &$4.500$ &$31$ &$2800$ \\
$0.200$, $0.220$          &$2592$ &$21$ &$1.500$ &$4.500$ &$31$ &$2800$ \\
$0.200$, $0.220$          &$3200$ &$21$ &$1.500$ &$4.500$ &$31$ &$2800$ \\
\hline
$0.500$                   & $512$ &$21$ &$0.212$ &$1.632$ &$30$ &$5964$ \\
$0.500$                   & $648$ &$24$ &$0.212$ &$1.632$ &$30$ &$5323$ \\
$0.500$                   & $800$ &$24$ &$0.212$ &$1.632$ &$30$ &$4859$ \\
$0.500$                   &$1152$ &$21$ &$0.373$ &$1.377$ &$21$ &$1140$ \\
\hline
Gauss                     & $512$ &$21$ &$0.212$ &$1.632$ &$30$ &$5000$ \\
Gauss                     & $648$ &$24$ &$0.212$ &$1.632$ &$30$ &$5149$ \\
Gauss                     & $800$ &$24$ &$0.212$ &$1.632$ &$30$ &$5053$ \\
Gauss                     &$1152$ &$24$ &$0.212$ &$1.632$ &$30$ &$1326$ \\
\hline
\hline
\end{tabular*}
\end{table}

\section{Results within the spin-glass sector}
\label{sec:results_sg}

\begin{figure}
\includegraphics[width=0.95\columnwidth]{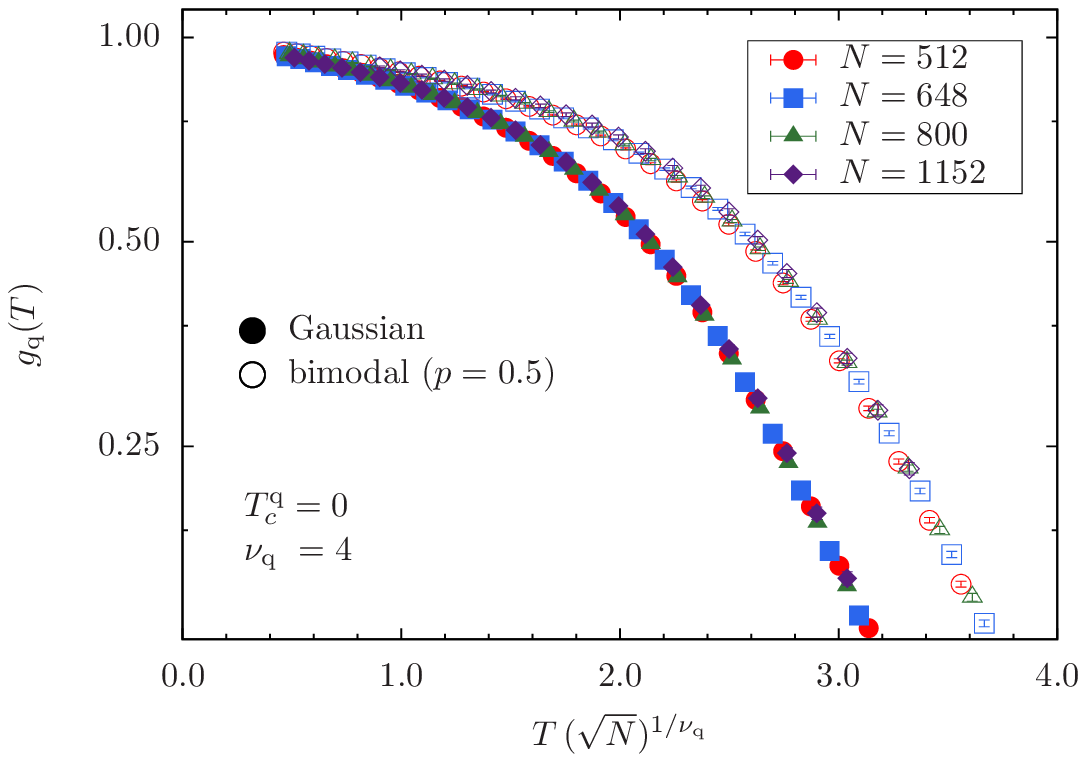}
\includegraphics[width=0.95\columnwidth]{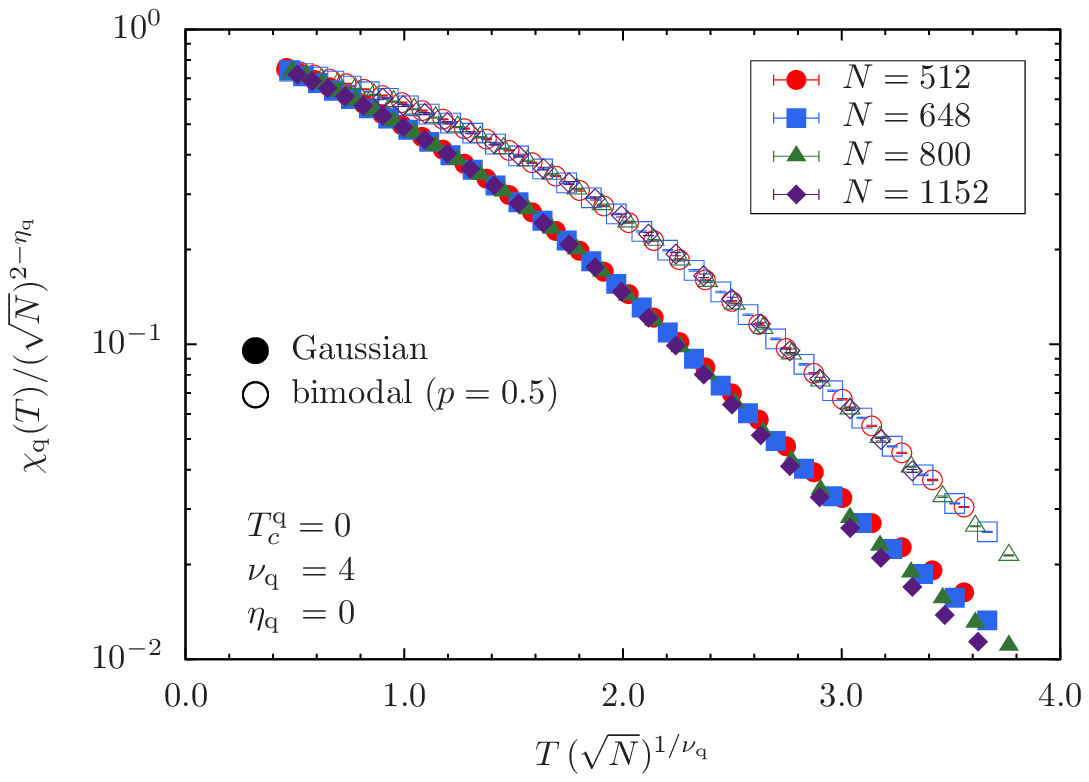}
\caption{
Top panel: Finite-size scaling of the Binder parameter $g_{\rm q}$ as a
function of $T(\sqrt{N})^{1/\nu_{\rm q}}$ for both Gaussian (full
symbols) and bimodal (open symbols) disorder on the Chimera topology.
The data scale very well for $T_c^{\rm q} = 0$ and $\nu_{\rm q} = 4$ in
both cases. Bottom panel: Finite-size scaling of the spin-glass
susceptibility $\chi_{\rm q}$ for both Gaussian and bimodal disorder.
Plotted are $\chi_{\rm q}/(\sqrt{N})^{2-\eta_{\rm q}}$ vs
$T(\sqrt{N})^{1/\nu_{\rm q}}$.  The data scale very well for $T_c^{\rm
q} = 0$, $\nu_{\rm q} = 4$, and $\eta_{\rm q} = 0$.  Note that there is
no universality violation. Error bars are smaller than the symbols.
}
\label{fig:sg}
\end{figure}

Figure \ref{fig:sg} shows a finite-size scaling analysis of the Binder
ratio (top panel) for both Gaussian disorder (full symbols) and bimodal
disorder with $p = 0.50$ (open symbols) in a semilogarithmic scale.  The
data scale extremely well, even far from the spin-glass transition
temperature. We find that for both cases, we have the same critical
parameters, namely \cite{comment:errors}
\begin{equation}
T_c^{\rm q}  = 0 \;\;\;\;\;\;\;\;\;\;\;\;\;\;
\nu_{\rm q}  \approx 4 \;\;\;\;\;\;\;\;\;\;\;\;\;\;
\eta_{\rm q} \approx 0 \, .
\label{eq:crit_q}
\end{equation}
Note that spin glasses on two-dimensional square lattices have $\nu
\approx 3.45$ \cite{katzgraber:04}; i.e., spin glasses on the Chimera
topology are close to two space dimensions.

Interestingly, the phase transition to a spin-glass phase only occurs at
zero temperature, despite the Chimera graph being nonplanar.
Furthermore, the divergence of the correlation length is rather violent,
with $\xi \sim T^{-4}$ for $T \to 0$. This suggests that the phase space
and correlations only build up close to $T = 0$ for a spin glass defined
on the Chimera topology. One of the potential advantages of a quantum
algorithm over a classical one lies in its ability to tunnel through
barriers. Classical algorithms must ``climb'' over these barriers
\cite{comment:alg}. The aforementioned results imply that the barriers
of a spin glass defined on a Chimera graph at nonzero temperature seem
to be of ``finite'' height, while for any Ising spin glass with a finite
transition temperature the barriers diverge below $T_c^{\rm q}$ for
decreasing temperature $T$ and increasing system size $N$. This could
offer one explanation for why quantum annealing machines, such as D-Wave
One and Two, cannot find a noticeable speedup over classical algorithms
such as vanilla simulated annealing \cite{kirkpatrick:83} on this class
of problems. Furthermore, a spin glass on a Chimera graph seems to order
in an almost ``discontinuous'' fashion at zero temperature with the
highly-connected blocks of eight spins behaving like a ``super spin'' on a
two-dimensional--like planar lattice. Once the individual blocks order,
the whole system suddenly orders. It is well known that quantum
annealing has problems when first-order transitions are present
\cite{young:08,amin:09,young:10,hen:11}; i.e., this could be the second
reason why quantum annealing machines do not seem to outperform simple
classical optimization methods on these problems.

\section{Results within the ferromagnetic sector}
\label{sec:results_fm}

\begin{figure}
\includegraphics[width=0.95\columnwidth]{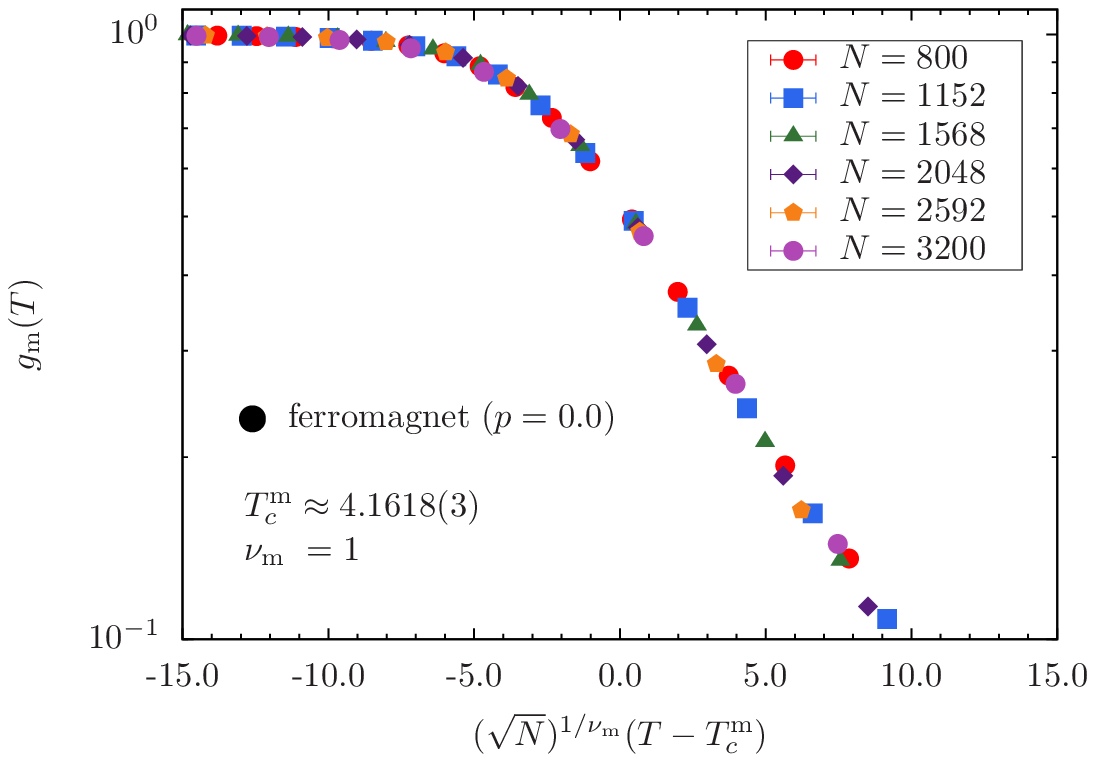}
\includegraphics[width=0.95\columnwidth]{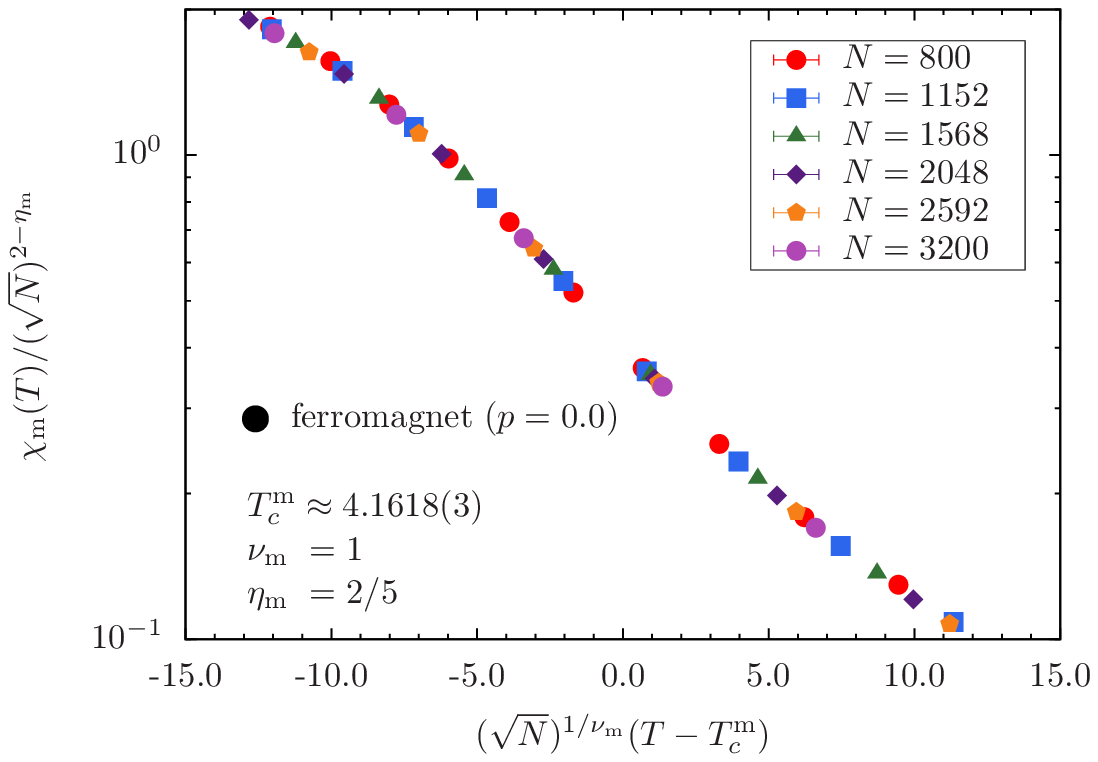}
\caption{
Top panel: Finite-size scaling of the ferromagnetic Binder parameter
$g_{\rm m}$ as a function of $(\sqrt{N})^{1/\nu_{\rm m}}(T - T_c^{\rm
m})$ and $p = 0$ (pure ferromagnet) on the Chimera topology. The data
scale very well for $T_c^{\rm m} = 4.1618(3)$ and $\nu_{\rm m} = 1$.
Bottom panel: Finite-size scaling of the ferromagnetic susceptibility
$\chi_{\rm m}$.  Plotted are $\chi_{\rm m}/(\sqrt{N})^{2-\eta_{\rm m}}$
vs $(\sqrt{N})^{1/\nu_{\rm m}}(T - T_c^{\rm m})$.  The data scale very
well for $T_c^{\rm m} \approx 4.1618(3)$, $\nu_{\rm m} = 1$, and
$\eta_{\rm m} = 2/5$.  Error bars are smaller than the symbols.
}
\label{fig:fm}
\end{figure}

For completeness, we also study the Ising ferromagnet on the Chimera
graph and compute the disorder $p$ (fraction of ferromagnetic bonds) vs
temperature $T$ phase diagram of the model. For no disorder, i.e., the
pure ferromagnet where $J_{ij} = 1$ $\forall i$, $j$ in
Eq.~\eqref{eq:ham}, an Ising model on the Chimera graph displays a
two-dimensional-Ising-model-like behavior. Figure \ref{fig:fm}, top
panel, shows a finite-size scaling analysis of the ferromagnetic Binder
ratio $g_{\rm m}$ as a function of the scaling variable
$(\sqrt{N})^{1/\nu_{\rm m}}(T - T_c^{\rm m})$. The data scale extremely
well for $T_c^{\rm m} = 4.1618(3)$ and $\nu_{\rm m} = 1$. Note that the
obtained value for the critical exponent $\nu_{\rm m}$ agrees with the
value for the two-dimensional Ising ferromagnet \cite{yeomans:92},
therefore corroborating our assumption that the system might behave
similarly to a two-dimensional superspin Ising model.

Figure \ref{fig:fm}, bottom panel, shows a finite-size scaling analysis
of the ferromagnetic susceptibility $\chi_{\rm m}/(\sqrt{N})^{2-\eta_{\rm
m}}$ as a function of $(\sqrt{N})^{1/\nu_{\rm m}}(T - T_c^{\rm m})$
using the estimate of the critical temperature determined from the
finite-size scaling of the Binder ratio. The data scale very well with
very small corrections to scaling using $\nu_{\rm m} = 1$ and $\eta_{\rm
m} = 2/5$. Note that the value of the critical exponent $\eta_{\rm m}$
is slightly larger, yet close to the exact value of the two-dimensional
Ising model ($\eta = 1/4$). Therefore, an Ising ferromagnet on the
Chimera graph and the two-dimensional Ising model do not share the same
universality class \cite{comment:exp}. In summary,
\begin{equation}
T_c^{\rm m}  = 4.1618(3) \;\;\;\;\;\;\;\;\;\;\;\;
\nu_{\rm m}  \approx 1 \;\;\;\;\;\;\;\;\;\;\;\;
\eta_{\rm m} \approx 2/5 \, .
\label{eq:crit_m}
\end{equation}

\begin{figure}
\includegraphics[width=0.95\columnwidth]{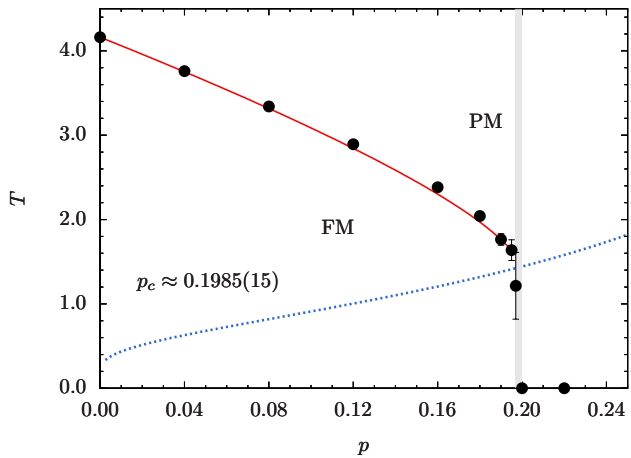}
\caption{
The disorder $p$ vs temperature $T = T_c^{\rm m}$ phase diagram for the
random-bond Ising model defined on the Chimera graph. The (blue) dotted
line represents the Nishimori line. The red (solid) line is a guide to
the eye. Under the solid curve, the system orders ferromagnetically (FM).
For $p \ge p_c \approx 0.1985(15)$, ferromagnetic order is lost and the
system is paramagnetic (PM). For $p \gtrsim p_c$ and $T = 0$, there is a
zero-temperature spin-glass state \cite{comment:reentrance}.
}
\label{fig:pd}
\end{figure}

Finally, we study the random-bond version of the Ising model on the
Chimera graph where a fraction $p$ of ferromagnetic bonds is
antiferromagnetic. We vary $p$ and compute the critical temperature of
the ferromagnetic phase. Figure \ref{fig:pd} shows the (critical)
temperature $T$ vs disorder $p$ phase diagram. The dotted (blue) line
represents the Nishimori condition \cite{nishimori:81}
$\exp(-2\beta)=p/(1-p)$. At the point where the phase boundary (solid
line) crosses the Nishimori line, i.e., for $p > p_c = 0.1985(15)$,
ferromagnetic order is lost.  This means that for any finite temperature
and $p \le p_c$, a random-bond Ising model on the Chimera graph is
essentially a disordered ferromagnet and is easily solved with a
conventional optimization algorithm. Therefore, to compare a quantum
adiabatic optimizer to any classical optimization method, strong enough
disorder is needed.

\begin{figure}
\includegraphics[width=0.90\columnwidth]{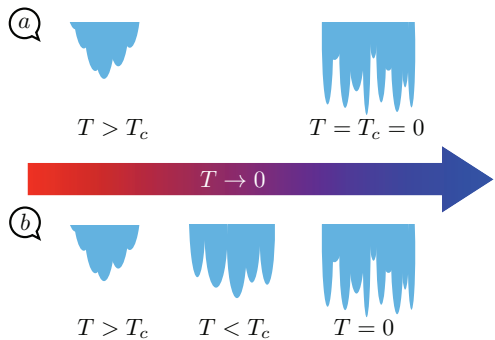}
\caption{
Sketch of the (coarse-grained) energy landscape for a system with a
zero-temperature transition [panel (a), top] and a finite-temperature
transition [panel (b), bottom] to a spin-glass state. For high enough
temperatures, i.e., above the critical temperature $T_c$, the energy
landscape is simple, with one clear minimum that dominates and some
``bumps along the way.'' For a system that has a finite-temperature
transition and for temperatures $T < T_c$, the energy landscape becomes
rough, with clear barriers that render any classical annealing schedule
inefficient, because the system can easily be trapped in a metastable
state for decreasing temperature. These barriers grow with increasing
system size $N$ and decreasing temperature $T$ until they form a rough
energy landscape for $T = 0$ (ground state of the system). For a system
where $T_c = 0$, such as spin glasses on the Chimera topology, the
energy landscape is typically simpler up until the ground state is
reached. This means that for such a system a classical annealing
schedule like simulated annealing should perform better in comparison to
quantum annealing which excels when the energy landscape exhibits
barriers \cite{comment:rsbvsdp}.
}
\label{fig:complexity}
\end{figure}

\section{Discussion}
\label{sec:discussion}

Figure \ref{fig:complexity} shows cartoons of the energy landscape for a
system with a zero-temperature transition [panel (a), top] and a
finite-temperature transition [panel (b), bottom] to a spin-glass state.
When the temperature is above the putative critical temperature $T_c$,
the energy landscape is typically simple with one dominant minimum.  For
a system that has a finite-temperature transition and for temperatures
$T < T_c$, the energy landscape becomes rough with barriers that grow
with decreasing temperature $T$ and increasing system size $N$ until the
ground state of the system is reached. However, for a system where $T_c
= 0$, such as spin glasses on the Chimera topology, the energy
landscape is likely much simpler until close to the ground state. This
means that for such a system, a classical annealing schedule should
perform well in comparison to quantum annealing which, in theory, excels
when the energy landscape has diverging barriers.  Note also that when
the system with bimodal disorder is in the ferromagnetic phase ($p <
p_c$), the energy landscape is also simpler and reminiscent of the
cartoon shown in the left-most panels of Fig.~\ref{fig:complexity}. This
means that benchmarking quantum annealing machines that operate at low,
but finite temperatures using spin glasses on a Chimera topology is
likely not the best approach.  Indeed, recent studies in a field
\cite{dash:13,saket:13} have shown that spin-glass instances can be
efficiently computed classically, albeit still scaling exponentially.
However, because there is likely no spin-glass state in a field
\cite{young:04,joerg:08a,katzgraber:09b,baity:13}, this is no surprise.
To truly discern if quantum annealing machines (defined on the Chimera
topology) display an advantage over classical annealing algorithms,
problems that display a finite-temperature transition and have a rough
energy landscape for a range of finite temperatures need to be embedded
in the machine's topology. Given the current hardware constraints, we
propose the following benchmarks:

{\em Three-dimensional cubic lattices}.--- The system has a
finite-temperature transition to a spin-glass state at $T_c \approx
0.96J$ for Gaussian disorder ($T_c \approx 1.1J$ for bimodal disorder
and $p = 0.5$) \cite{katzgraber:06}. We estimate that a Chimera graph of
$2048$ qubits could be used to embed a relatively modest 3D system of
$5^3=125$ spins with periodic boundary conditions and one of size
$8^3=512$ with free boundary conditions.  Note that current state-of-the
art classical optimization algorithms \cite{hartmann:01,hartmann:04} can
estimate ground states to high accuracy of up to approximately $14^3 =
2744$ spins.

{\em Viana-Bray model}.--- The Ising spin glass is defined on a random
graph with average connectivity $k$ \cite{viana:85,katzgraber:01}.  For
any $k > 2$, the system has a finite-temperature phase transition into a
spin-glass state. To simplify the embedding, a random graph with
Gaussian disorder and $k= 3$ could be studied, where $T_c \approx
0.748J$. However, to be able to embed the long-range connections between
the spins, we estimate that ${\cal O}(N^2)$ qubits might be needed to
embed a system with $N$ spins like in the mean-field
Sherrington-Kirkpatrick model \cite{sherrington:75,choi:08,choi:11}.

{\em Rescaled Chimera systems}.--- It is plausible that if the ratio of
interactions within the cells and between the cells changes proportional
to the system size, a mean-field-like finite-temperature spin-glass
transition might emerge. For example, the random intercell interactions
could be rescaled with $J_{ij} \to J_{ij}/f(N)$ [where $f(N)$ is a
nondecreasing function of the system size] while leaving the random
intracell interactions untouched. We attempted to weaken the effects of
the tightly bound intercell spin clusters in the Chimera graph by
setting the spin-spin interactions to $1/4$ of all intracell
interactions (on average). Although universality considerations would
suggest that $T_c$ should still be zero when $f(N) = 1/4$ $\forall N$,
our data for systems up to approximately $3200$ spins suggest $T_c
\approx 0.6(2)$. We do emphasize, however, that corrections to scaling
are huge and a study with far larger systems might be needed.

One could, in principle, also embed a two-dimensional Ising spin glass
on a square lattice, where it was first shown by Santoro {\em et
al.}~that quantum annealing might display an advantage over classical
annealing via simulations at very low temperatures \cite{santoro:02}.
However, ground states of two-dimensional Ising spin glasses can be
computed in polynomial time, and the low-temperature behavior of this
system is known to be unusual and still controversial \cite{thomas:11c}.
As such, this might not be a robust and well-controlled benchmark,
especially because the D-Wave machines operate at temperatures
considerably higher than in the aforementioned study by Santoro {\em et
al.}~\cite{santoro:02}.

The previous examples also illustrate a limitation of the Chimera
topology: To embed many systems, a large overhead of quantum bits in the
machine to simulated physical bits is needed. This is particularly the
case because no long-range connections between the spins are present---a
feature that should be included in future chip designs.  Finally, at
this point it is unclear if the critical behavior of an {\em embedded}
system is the same as the critical behavior of the {\em actual}
classical system. This is of utmost importance if one wants to use
programmable quantum annealing machines as quantum simulators.

\section{Conclusions}
\label{sec:conclusions}

Although Barahona \cite{barahona:82} has rigorously shown that spin
glasses defined on graphs like the Chimera topology are worst-case
NP-hard types, the Chimera spin glass of the type used so far to compare
quantum to classical annealers represents a hard, but typically easier
optimization problem. Because such a spin glass on the Chimera topology
only orders at zero temperature, classical thermal annealers will
typically be able to efficiently estimate ground states for the system.
The performance of these classical algorithms would considerably
deteriorate if the problem to be optimized exhibits a finite-temperature
transition below which energy barriers diverge with decreasing
temperature and increasing system size. To be able to show that quantum
annealing machines based on the Chimera topology outperform classical
annealing schedules, nontrivial embeddings in higher space dimensions or
with long-range interactions, as outlined above, are needed.
Furthermore, at this point it is unclear how the overhead of the
embedding scales with the size of the system and if the embedded system
via edge contraction shares the same universality class as the true
problem to be emulated---especially when simulated on the actual D-Wave
hardware \cite{comment:noise}. The latter open questions are the subject
of current research and we conclude by emphasizing that the design of
the hardware topology in quantum annealing machines is of crucial
importance.

\begin{acknowledgments}

We would like to thank F.~Barahona, P.~Bunyk, J.~Machta, R.~B.~McDonald,
C.~Moore, M.~A.~Moore, H.~Nishimori, T.~F.~R{\o}nnow, C.~K.~Thomas,
M.~Troyer, A.~P.~Young, and I.~Zintchenko for fruitful discussions. In
particular, we would like to thank R.~B.~McDonald for porting the Python
Chimera graph generator to \texttt{C}.  H.G.K.~acknowledges support from
the NSF (Grant No.~DMR-1151387) and would also like to thank
Bruichladdich Dist.~for providing the initial inspiration for this
project, as well as the Royal Street Courtyard Inn in New Orleans for
their hospitality during the final stages of this manuscript. Finally,
we would like to thank the Texas Advanced Computing Center (TACC) at The
University of Texas at Austin for providing HPC resources (Lonestar and
Stampede Clusters) and Texas A\&M University for access to their Eos
cluster.

\end{acknowledgments}

\bibliography{refs,comments}

\begin{thebibliography}{64}
\bibitem[{com({\natexlab{c}})}]{comment:tail}
\bibinfo{note}{It is very important to study the distributions of times and not
  just averages. While the worst-case scenario is that there is at least one
  instance that cannot be treated efficiently, our claim is that, given the
  energy landscape, one can expect that the distribution of time scales to
  probe different ground-state instances likely does not have as fat a tail as,
  for example, a three-dimensional system with a finite transition temperature.
  In other words, the number of hard instances is small. However, the vast
  majority of instances can be treated rather efficiently with simple
  optimization methods, such as simulated annealing. Finally, all classical
  optimizers, as well as the D-Wave machine operate at nonzero temperatures.
  Because the system orders only at zero temperature, the path to low-lying
  states will be typically easier to reach than if the system had a finite
  transition temperature where barriers would grow in a more pronounced
  fashion.}
\bibitem[{com({\natexlab{d}})}]{comment:errors}
\bibinfo{note}{Note that we do not list error bars for the statistical
  estimates of the critical parameters. This is because the obtained results
  agree to high precision with the exact (often integer) values quoted. We also
  emphasize that the exact values of the critical exponents do not affect our
  main results. In particular, we feel it is unlikely that $T_c > 0$ for spin
  glasses on the Chimera topology.}
\bibitem[{com({\natexlab{e}})}]{comment:alg}
\bibinfo{note}{Note that simulated annealing---by design---is unable to climb
  over barriers because temperature is monotonically quenched to zero. However,
  more efficient algorithms like exchange Monte Carlo \cite{hukushima:96} are
  able to ``climb'' over barriers.}
\bibitem[{com({\natexlab{f}})}]{comment:exp}
\bibinfo{note}{To fully determine a universality class two critical exponents
  are needed (here $\eta$ and $\nu$).}
\bibitem[{com({\natexlab{g}})}]{comment:reentrance}
\bibinfo{note}{Note that the disorder vs temperature phase diagrams are
  typically reentrant for disordered ferromagnets. Therefore, it is safe to
  assume that a spin-glass phase at zero temperature might exist for values of
  $p$ slightly smaller than $p_c$. For examples of reentrant phase diagrams for
  very different model systems, see Refs.~\cite{katzgraber:09c}
  and\cite{thomas:11d}.}
\bibitem[{com({\natexlab{h}})}]{comment:rsbvsdp}
\bibinfo{note}{Note that the statements made are independent of which theory
  (droplet \cite{fisher:86,fisher:87,fisher:88,bray:86} vs replica symmetry
  breaking \cite{parisi:79,parisi:80,parisi:83,mezard:87}) describes the
  low-temperature state of spin glasses best. For the numerically accessible
  system sizes $N$, there are features in the overlap distribution like in the
  mean-field model. The replica symmetry breaking versus droplet picture
  question, however, only affects results in the thermodynamic limit and not in
  the finite systems probed in simulations. In the replica symmetry breaking
  picture for the mean-field model, the number of states increases by around
  $N^{1/6}$ \cite{aspelmeier:08}. For the droplet model, there is only one
  dominant state in the thermodynamic limit. Furthermore, in the mean-field
  limit, barriers should increase by around $N^{1/3}$, whereas for a
  short-range model the growth is to date unclear. Therefore, at low, but
  finite temperature, the replica symmetry breaking vs droplet picture
  controversy should not play a role if $N < \infty$. In both pictures (below a
  finite $T_c$), barriers in the energy landscape appear that grow with
  increasing system size and decreasing temperature (at least for any
  numerically accessible system sizes). We emphasize that what matters is that
  barriers grow for any temperature $T \le T_c$.}
\bibitem[{com({\natexlab{i}})}]{comment:noise}
\bibinfo{note}{These considerations do not take hardware noise effects on the
  simulated universality classes into account -- a problem that should also be
  studied in detail.}
\end{thebibliography}

\end{document}